\documentstyle[amssymb,preprint,aps]{revtex}

\draft

\begin{document}
\title{Evidence for Internal Field in Graphite: a Conduction Electron Spin
Resonance Study}
\author{M.S. Sercheli, Y. Kopelevich, R. Ricardo da Silva*, J.H.S. Torres, and C.
Rettori,}
\address{Instituto de F\'{i}sica ``Gleb Wataghin'', UNICAMP,13083-970, Campinas-SP,\\
Brazil.}
\date{\today }
\maketitle

\begin{abstract}
We report conduction electron spin resonance measurements performed on
highly oriented pyrolitic graphite samples between $10$ K and $300$ K using
S ($\nu $ $=4$ GHz), X ($\nu $ $=9.4$ GHz), and Q ($\nu $ $=34$.4 GHz)
microwave bands for the external $dc$-magnetic field applied parallel ($%
H\parallel c$) and perpendicular ($H\perp c$) to the sample hexagonal $%
c-axis $. The results obtained in the $H\parallel c$ geometry are
interpreted in terms of the presence of an effective internal
ferromagnetic-like field, $H_{int}^{eff}$($T,H$), that increases as the
temperature decreases and the applied $dc$-magnetic field increases. We
associate the occurrence of the $H_{int}^{eff}$($T,H$) with the
field-induced metal-insulator transition in graphite and discuss its origin
in the light of relevant theoretical models.
\end{abstract}


Recent experiments revealed a number of unexpected phenomena in graphite
such as a linear energy dependence of the quasiparticle damping, \cite{Xu}
high-$T$ superconducting and ferromagnetic (FM) correlations,\cite
{Kopelevich,Y. Kopelevich} and magnetic-field-induced metal-insulator
transition (MIT).\cite{Kopelevich,Kempa} These properties could be
understood taking into account electron-electron interactions within the
graphite layers.\cite{Gonzales} A possible proximity of graphite monolayer
to the excitonic instability was an alternative explanation for both, the
linear energy dependence of the quasiparticle damping and the weak FM\cite
{Khveshchenko}. These observations bring graphite to the class of materials
which attract a broad scientific interest, including high-$T_{c}$
superconducting (HTS) cuprates, in which a strong Coulomb repulsion between
carriers is the key issue,\cite{Anderson} hexaborides demonstrating a high-$%
T $ weak FM,\cite{Young} and most recently discovered superconducting MgB$%
_{2}$, \cite{Nagamatsu} a material isoelectronic to graphite. \cite{Baskaran}

It has been proved long ago \cite{Dresselhaus} that the electron spin
resonance (ESR) is a powerful experimental technique that allows to probe
directly the properties of conduction electrons (holes) in graphite.
However, the observed $g$-factor anisotropy and its temperature dependence
remains unexplained by existing theories which consider{\bf \ }the
spin-orbit coupling mechanism{\bf ,} \cite{Dresselhaus} indicating the
necessity of further experimental and theoretical work.

In this letter we report the experimental results which provide a fresh
insight on the conduction ESR (CESR) in highly oriented pyrolitic graphite
(HOPG) samples. In particular, we show that the results can be interpreted
in terms of an effective internal FM-like field, $H_{int}^{eff}(T,H)$,
induced by the external $dc$-magnetic field, $H_{ext}=H$, applied along the
sample hexagonal $c-axis$. The results obtained in this work indicate that
the occurrence of $H_{int}^{eff}(T,H)$ intimately couples to the
magnetic-field-induced MIT.

CESR measurements performed on three HOPG samples, synthesized at the
Research Institute ''Graphite'' (Moscow) \cite{Kopelevich,Y.
Kopelevich,Kempa} and the Union Carbide Co. were carried out in a
temperature interval from $10$ K to $300$ K by means of a Bruker ELEXSYS-CW
spectrometer operating at three microwave frequency bands, S ($\nu =4$ GHz),
X ($\nu $ $=9.4$ GHz), and Q ($\nu $ $=34.4$ GHz). The measurements with
mutually perpendicular microwave and $dc$-magnetic fields were performed in
both $H\parallel c$ and $H\perp c$ configurations. In order to warrant a
maximal microwave penetration and to avoid a line-shape change caused by
anisotropic skin depth effects, \cite{Walmsley} the microwave field was
always kept perpendicular to the sample $c-axis$. Similar results were
obtained for all graphite samples studied in this work. Complementary
low-frequency ($\nu =1$ Hz) standard four-probe basal-plane resistance $%
R(T,H)$ and $dc$-Hall effect measurements were performed for $H\parallel c$
using a Quantum Design PPMS-$9$ T and Janis-$9$ T magnet systems.

The obtained low field in-plane carrier mobility $\sim 7\cdot 10^{5}-2\cdot
10^{6}$ cm$^{2}$/V$\cdot $s ($T=4.2$ K) indicates the high quality of our
samples. \cite{Dresselhaus} The high sample quality has also been confirmed
by means of Shubnikov-de Hass oscillations measurements, where, e.g., two
resistance minima at $H=6.7$ T and $H=7.5$ T, corresponding to the spin
splitting of the $n=1$ Landau level for electrons,\cite{Iye,Timp,Y.Iye}
could clearly be resolved.

In all our CESR measurements a Dysonian line-shape with $A/B$ asymmetry
ratio between $4$ and $9$ was observed.\cite{Dyson} The CESR parameters
(resonance field and line-width) were obtained assuming an admixture of
absorption/dispersion of a Lorentzian line. \cite{Pake} The integrated
absorption part of the resonance was found to be $T$-independent within the
accuracy of the measurements, confirming that the ESR originates from
itinerant carriers. \cite{Abragam,Taylor}

Figure 1 (a, b, c) presents the $T$-dependencies of the resonance field $%
H_{R}(T)$ for one of our studied samples (HOPG-3, see Ref. [4]) in $%
H\parallel c$ and $H\perp c$ geometry using S, X, and Q bands, respectively.
As can be seen from Fig. 1, $H_{R}$ obtained for $H\perp c$ is $T$%
-independent within the experimental resolution, whereas $H_{R}(T)$ measured
for $H\parallel c$ is a decreasing function of $T$. Note, that $%
H_{R\parallel }-H_{R\perp }<0$ which can be re-written in terms of the Lande 
$g$-factor shift $\Delta g=g_{\parallel }-g_{\perp }>0$. This is a well
known experimental fact being, however, in conflict with the theoretical
predictions of $\Delta g<0$ based on the spin-orbit interaction. \cite
{Dresselhaus} Figure 2 shows $T$-dependencies of the difference $\Delta
H_{R}=H_{R\perp }-H_{R\parallel }$ extracted from the data of Fig. 1. These
data demonstrate that $\Delta H_{R}(T,H)$ increases as the temperature
decreases and the measuring field (frequency) increases. The inset in Fig. 2
depicts the behavior of $g_{\parallel }$ also extracted from the data of
Fig. 1. As can be seen from this plot, the $g_{\parallel }(T,H)$ is both $T$%
- and $H$-dependent. Note a rather complex and non-systematic behavior of $%
g_{\parallel }(T,H)$: while $g_{\parallel }(T,H)$ obtained from the S band
data coincides with that obtained from Q band for $T\gtrsim 70$\ K, it
coincides with the $g_{\parallel }(T,H)$ measured with X band for $T<$\ $70$%
\ K, and all three bands give the same value of the $g$-factor at $T=300$\ K
(see inset in Fig. 2){\bf . }On the other hand, a comparative analysis of
the CESR and transport measurements revealed an intriguing correlation
between $\Delta H_{R}(T,H)$ and the basal-plane resistance, $R(T,H)$,
suggesting that the description of the results presented in Fig. 1 in terms
of $\Delta H_{R}(T,H)$ is more appropriate than that in terms of $%
g_{\parallel }(T,H)$. It is shown in Fig. 2 that $\Delta H_{R}(T,H)$ can be
well approximated by the empirical equation:

\begin{equation}
\Delta H_{R}(T,H)=H_{0}ln[R(T,H)/R_{0}(H)]
\end{equation}
where $H_{0}$ and $R_{0}(H)$ are fitting parameters. Both, $\Delta
H_{R}(T,H) $ and $R(T,H)$ decrease approaching the transition from
insulating-like ($dR/dT<0$) to the metallic-like ($dR/dT>0$) resistance
behavior,\cite{Kempa} implying that the appearance of the $\Delta
H_{R}(T,H)>0$ is essentially coupled to the field-induced MIT.

It is already established that the CESR results of Fig. 1 are characteristic
of both HOPG and single crystalline graphite.\cite{Dresselhaus} Because this
and in order to be more specific on our claim regarding the relationship
between $\Delta H_{R}(T,H)$ and MIT, we present in Fig. 3 (a, b) the $R(T,H)$
data obtained for HOPG (Union Carbide Co.) and Kish single crystalline
graphite samples. The data of Fig. 3 (a, b) provide an unambiguous evidence
that the field-induced MIT is an intrinsic property of graphite. We note
that the low temperature leveling-off of the insulating-like $R(T,H)$ is
associated with the reentrant transition to the metallic state, \cite
{Kopelevich} which will be not discussed in the present work. A careful
analysis of the results given in Fig. 3 (a, b) revealed that the
insulating-like resistance $R(T,H)$ develops below a well defined
field-dependent temperature $T_{min}(H)$ which is an increasing function of $%
H$. This is exemplified for the Kish graphite in Fig. 4 where the reduced
resistance $R(T)/R(T_{min})$ vs$.$ $T$ is plotted. The inset in Fig. 4 shows 
$T_{min}(H)$ which can be best described (dashed curve) by the formula:

\begin{equation}
T_{min}(H)=A(H-H_{c})^{1/2}  \label{eq. 1}
\end{equation}
with $A$ and $H_{c}$ being the fitting parameters. For $H>1$ kOe, no minimum
in $R(T)$ has been found in the studied temperature interval.

In attempts to understand the origin of $\Delta H_{R}(T,H)>0$, which
suggests the existence of an effective internal FM-like field $%
H_{int}^{eff}(T,H)$, as well as its coupling to the MIT we note that the
characteristic feature of the band structure of a single graphene layer is
that there are two isolated points in the first Brillouin zone where the
band dispersion is linear $E$({\bf k}) = $\hbar $v$\mid ${\bf k}$\mid $(v = v%
$_{F}$ $\sim 10^{6}$ m/s is the Fermi velocity), so that the electronic
states can be described in terms of the Dirac equations in two dimensions. 
\cite{Gonzales,Khveshchenko,Balents} Besides, an opening of the gap in the
spectrum of Dirac fermions by an applied magnetic field (the so-called
''magnetic catalysis'' phenomenon), accompanied by the time-reversal
symmetry breaking, has been discussed in a number of recent theoretical
works (see e.g., \cite{Laughlin,Ferrer}, and references therein). The
equation for the field-induced transition temperature $T_{c}(H)$ to the
gapped state is given by:

\begin{equation}
k_{B}T_{c}(H)=C\hbar \text{v}(2e/\hbar )^{1/2}(H-H_{c})^{1/2},
\end{equation}
which is similar to the Eq. (2) derived from the experiment. Here $C$ is the
constant, $\Delta _{0}$ = $\hbar $v($2e/\hbar $)$^{1/2}$($H-H_{c}$)$^{1/2}$
is the induced zero-temperature gap, and $H_{c}$ is the critical field.
According to the theory, the development of a magnetic moment proportional
to the $\Delta (T,H)$ is also expected. In the context of the present work
this would imply $H_{int}^{eff}(T,H)\sim \Delta (T,H)$ which can account for
the correlation between $H_{int}^{eff}(T,H)$ and $R(T,H)$, see Fig. 2.
Whereas in Refs. [18, 19] the gap occurrence is related to the development
of an additional superconducting order parameter in HTS, the gap in the case
of graphite is associated with the insulating-like state. However, as
emphasized by Laughlin \cite{Laughlin} the theory [Eq. (3)] considers energy
scales only. In the light of theoretical results of Ref. [6], it is possible
that the here observed phenomenon is related to the field-driven
metal-excitonic insulator transition.

While the analysis of the experimental results in terms of an effective
internal FM-like field occurrence appears to be self-consistent, we cannot
exclude a complex dependence of the $g$-factor on the applied magnetic field
and temperature (see inset in Fig.2) as an alternative approach. However a
lack of a proper theory for both magnetic field and temperature dependence 
\cite{Dresselhaus} of the $g$-factor prevents us to speculate on this matter.

Finally, we present in Fig. 5 the CESR half line-width, $\Delta H_{1/2}(T)$,
measured at various frequencies ($\nu $ $=4$ GHz, $\nu $ $=9.4$ GHz, and $%
\nu =34.4$ GHz). Due to an intrinsic $mosaic$ effect in HOPG the line-width
for $H\parallel c$ is always broader than for $H\perp c$. As can be observed
in Fig. 5, $\Delta H_{1/2}(T)$ increases with temperature decreasing, in
agreement with previous studies.\cite{Dresselhaus} The novel result,
however, is that $\Delta H_{1/2}(T)$ is essentially field
(frequency)-independent. Then, the induced effective internal FM-like field
should be highly uniform. In principle, various physical processes such as a
distribution of $g$-factor values, motional narrowing, many-body exchange,
and relaxation effects can contribute to $\Delta H_{1/2}(T)$,\cite
{Dresselhaus,Lubzens} so that a further work is needed to provide a
plausible explanation for $\Delta H_{1/2}(T)$. For example, pulsed CESR
experiments where the spin-spin and spin-lattice relaxation times can be
measured, are highly desirable.

In summary, CESR measurements performed on HOPG samples using S ($\nu =4$
GHz), X ($\nu $ $=9.4$ GHz), and Q ($\nu =34.4$ GHz) microwave bands
revealed a complex behavior of the $g$-factor as a function of the applied
magnetic field which cannot be understood within a framework of the existing
theories. Alternatively, we show that the results can be self-consistently
described in terms of the occurrence of an effective internal FM-like field
induced by the applied $dc$-magnetic field ($H\parallel c-axis$). It is also
found that the occurrence of the internal field is intimately coupled to\
the magnetic field-induced MIT. We argue that the magnetic field-induced gap
in the energy spectrum of Dirac fermions in graphite can be responsible for
the here observed phenomena.

{\it \bigskip }

This work is supported by FAPESP, CNPq, and CAPES Brazilian science
agencies. We thank V.V. Lemanov and P. Scharff for supplying the graphite
samples used in the present study, and D.V. Khveshchenko for discussions.

* Also at FAENQUIL, 12600-000, Lorena, SP, Brazil.

\begin{figure}[tbp]
\caption{{}$T$-dependencies of the resonance field, $H_{R}(T)$, measured in
the HOPG-3 sample for $H$ $\Vert $ $c$ (squares) and $H$ $\bot $ $c$
(triangles) geometries using S ($\protect\nu =4$ GHz), X ($\protect\nu =9.4$
GHz), and Q($\protect\nu =34.4$ GHz) microwave bands. The indicated $g$%
-factors correspond to the values at room-$T$. The continuous lines are
guide for the eye.}
\label{Figure 1}
\end{figure}

\begin{figure}[tbp]
\caption{$\Delta H_{R}(T)$ obtained from the data of Fig. 1 (solid symbols).
Open symbols correspond to $\Delta H_{R}(T)$ calculated from Eq. (1) using
basal-plane resistance data ($H\parallel c$) obtained for the HOPG-3 sample
with $H=1.34$ kOe, ($\bigtriangleup $), $H=3.14$ kOe ($\Box $), and $H=11.5$
kOe ($\circ $) and fitting parameters $H_{0}=150$ Oe, $R_{0}=0.014$ $\Omega $%
; $H_{0}=150$ Oe, $R_{0}=0.01$ $\Omega $; $H_{0}=150$ Oe, $R_{0}=0.004$ $%
\Omega $, respectively. Inset depicts $g_{\parallel }(T,H$) extracted from
the $H_{R}(T,H)$ data given in Fig. 1. The continuous lines are guide for
the eye.}
\end{figure}
\begin{figure}[tbp]
\caption{Basal-plane resistance, $R(T,H)$, measured at various magnetic
fields (in kOe) applied parallel to the sample $c-axis$ in (a) HOPG (Union
Carbide Co.) and (b) single crystalline Kish graphite. }
\end{figure}
\begin{figure}[tbp]
\caption{ Reduced basal-plane resistance $R(T)/R(T_{min})$ measured in Kish
graphite single crystal with $H=0.04$ T ($\Box $), $H=0.05$ T ($\circ $), $%
H=0.075$ T ($\triangle $), and $H=0.1$ T ($\triangledown $) applied parallel
to the $c-axis$, where $T_{min}(H)$, denoted by arrows, is the
field-dependent temperature separating insulating-like ($T<T_{min}$) and
metallic-like ($T>T_{min}$) resistance behavior. The continuous lines are
guide for the eye. Inset shows $T_{min}(H)$; the dashed line is obtained
from Eq. (2) with the fitting parameters $A=350$ K/T$^{1/2}$ and $H_{c}=0.038
$ T. }
\end{figure}

\begin{figure}[tbp]
\caption{ The CESR half line-width, $\Delta H_{1/2}(T)$, measured in the
HOPG-3 sample for both $H\parallel c$ and $H\perp c$ geometries using S ($%
\protect\nu =4$ GHz), X ($\protect\nu =9.4$ GHz), and Q($\protect\nu =34.4$
GHz) microwave bands. The dotted lines are guide for the eye.}
\end{figure}

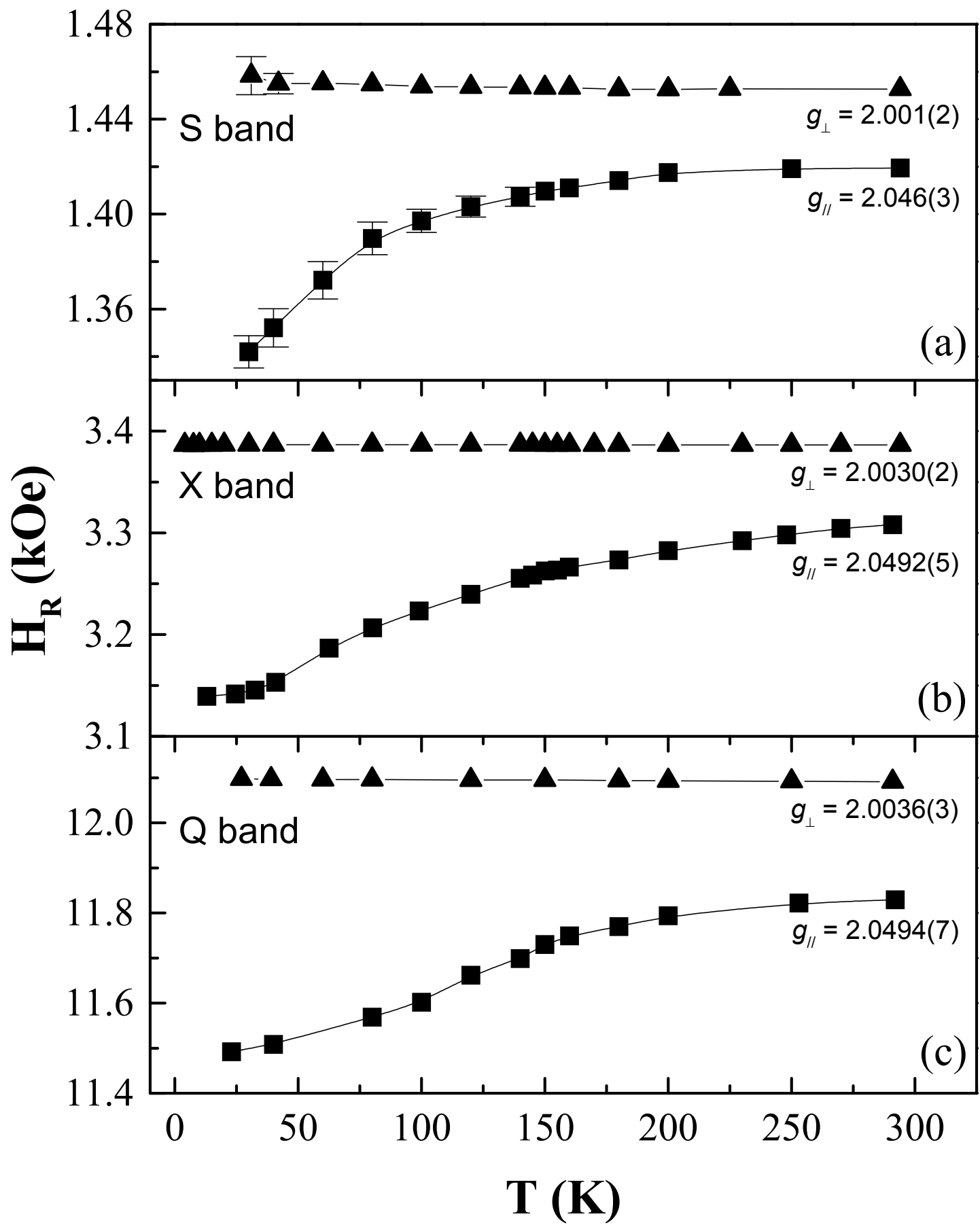
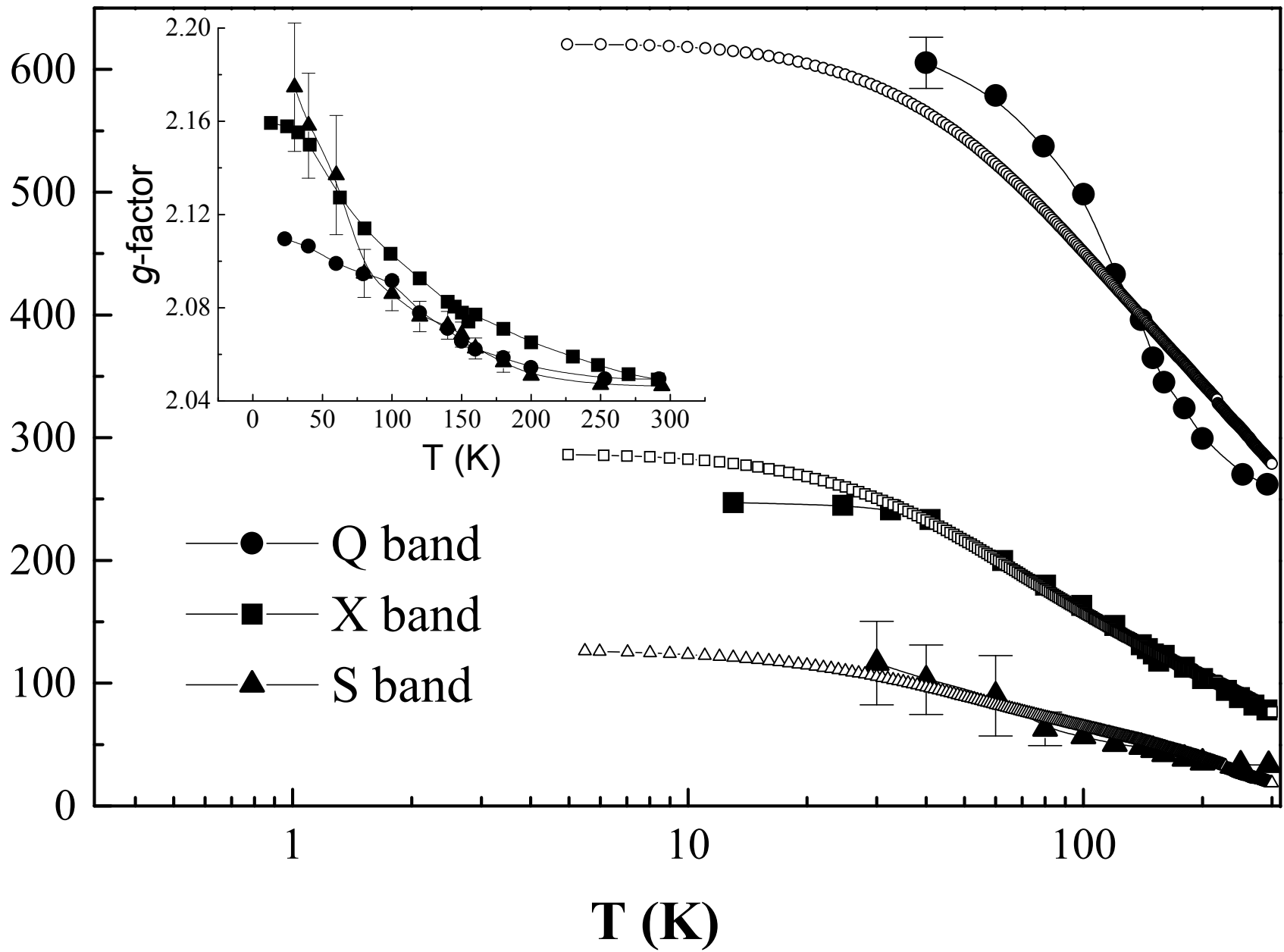
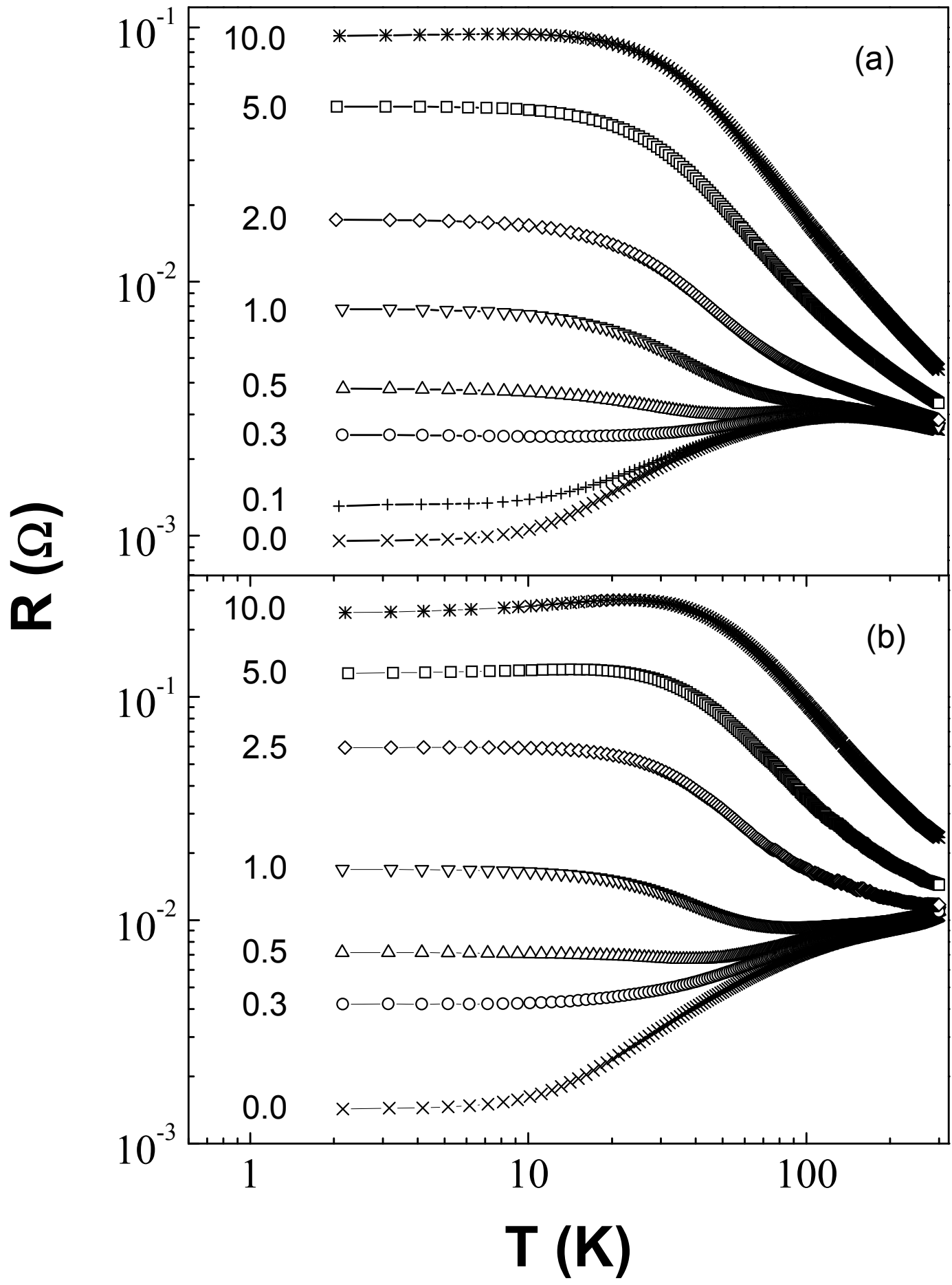
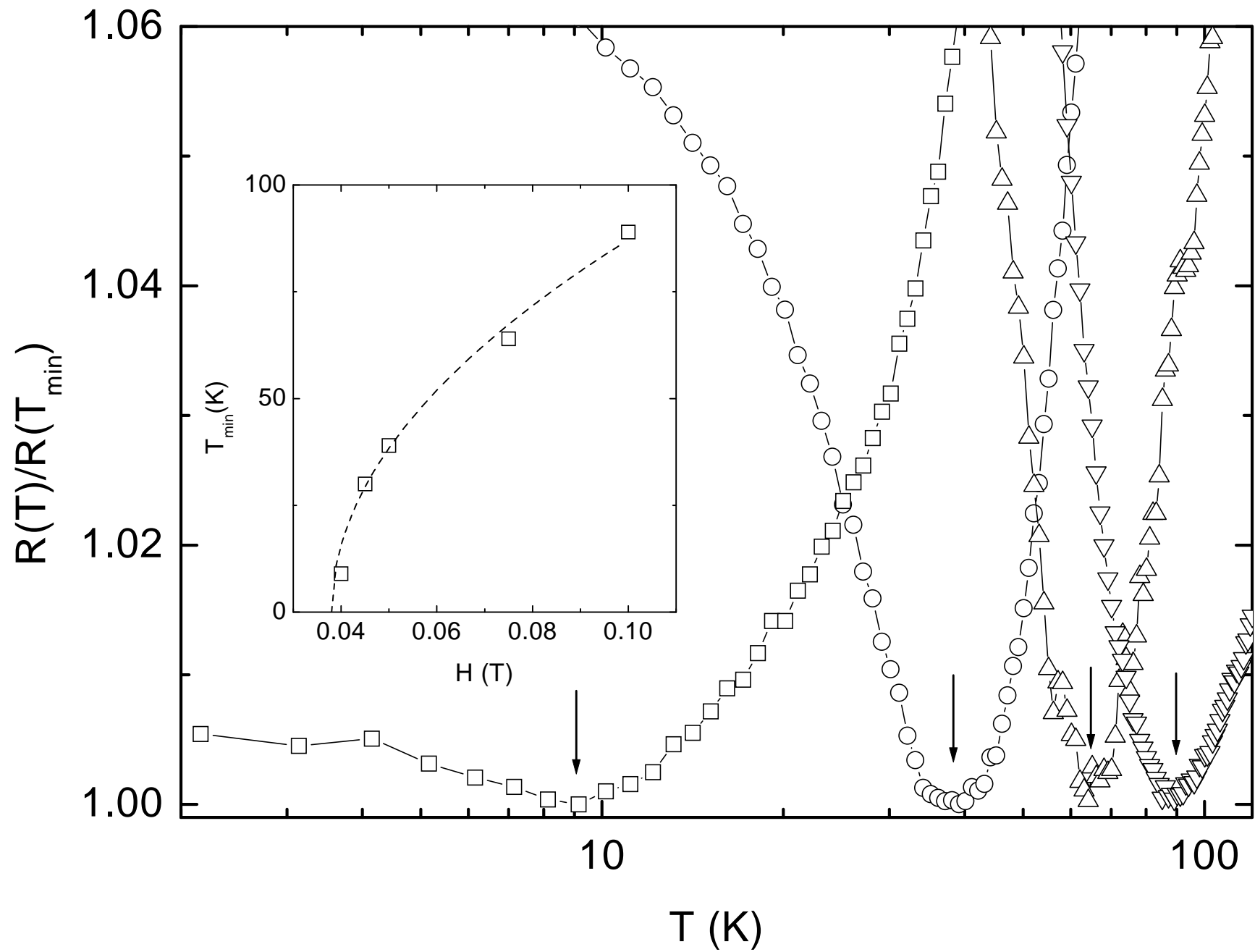
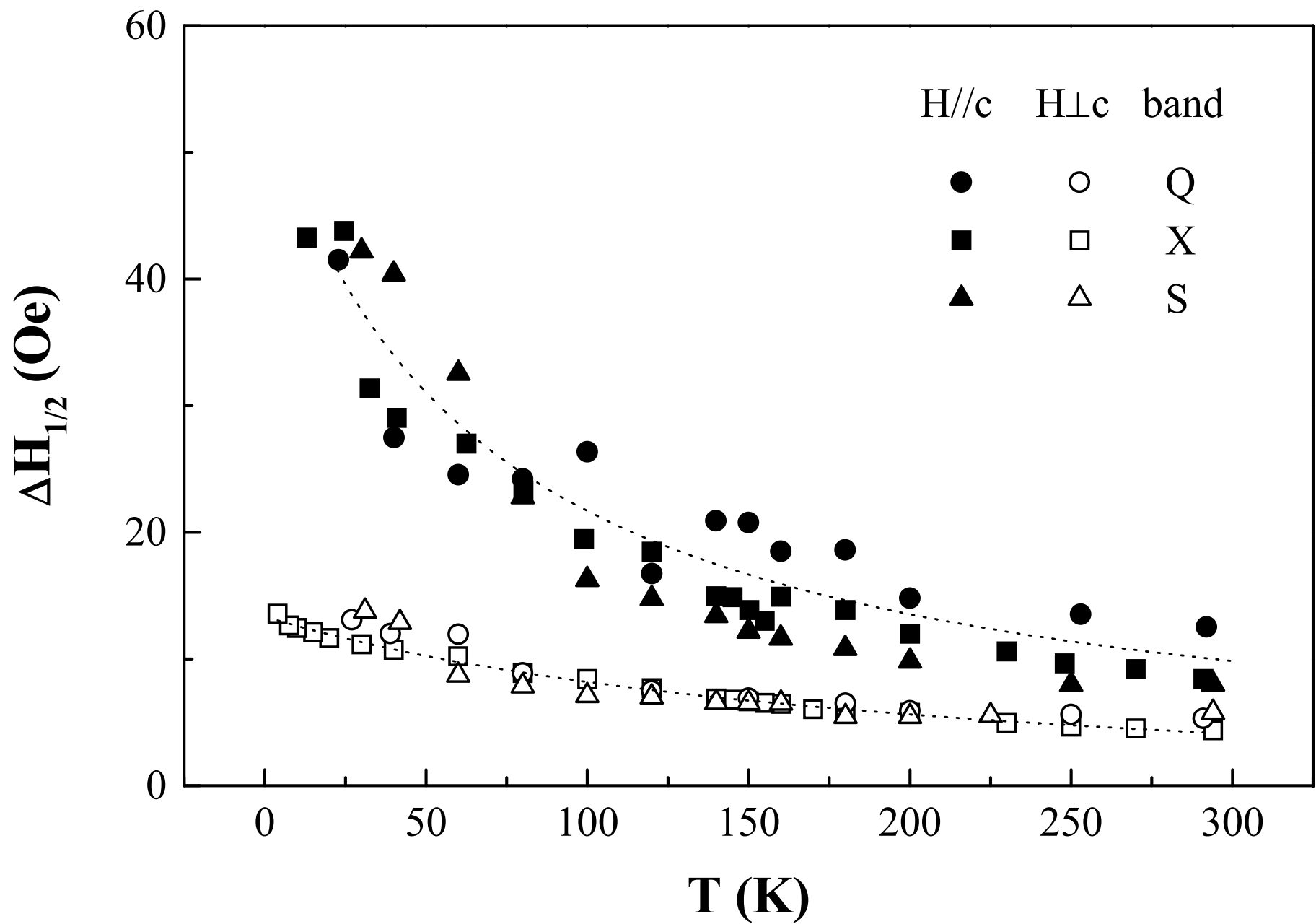
\end{document}